\documentclass[prl,twocolumn,superscriptaddress]{revtex4}
\usepackage{epsfig,amsmath,amssymb,graphicx,color,calc}

   \let\g=\gamma  \let\d=\delta \let\e=\varepsilon
     \let\l=\lambda
      \let\o=\omega  \let\x=\xi     \let\p=\pi   
\let\s=\sigma \let\t=\tau    
   
\let\G=\Gamma \let\D=\Delta   
 \let\Si=\Sigma     \let\P=\Psi 
  \let\r=\rho \let\th=\theta
\let\io=\infty \let\O=\Omega

\def\MM{{\cal M}}

\def\vx{\vec{x}} \def\vy{\vec{y}} \def\vz{\vec{z}} \def\vk{\vec{k}} \def\vr{\vec{r}}  \def\vq{\vec{q}}      \def\grad{\vec{\nabla}}

 \def\olP{{\overline{\P}}} \def\ola{{\overline{a}}}

\def\to{\rightarrow}
\def\la{\left\langle}
\def\ra{\right\rangle}

\begin{document}

\title{Field theoretic formulation of a mode-coupling equation for colloids}

\author{Hugo Jacquin}
\author{Fr\'ed\'eric van Wijland}
\affiliation{Laboratoire Mati\`ere et Syst\`emes Complexes, UMR CNRS 7057,
Universit\'e Paris Diderot -- Paris 7, 10 rue Alice Domon et L\'eonie 
Duquet, 75205 Paris cedex 13, France}

\date{\today}

\begin{abstract}
The only available quantitative description of the slowing down of the dynamics upon 
approaching the glass transition has been, so far, the mode-coupling theory, developed 
in the 80's by G\"otze and collaborators. 
The standard derivation of this theory
does not result from a systematic expansion. We present a field theoretic formulation that 
arrives at very similar mode-coupling equation but which is based on a variational principle 
and on a controlled expansion in a small dimensioneless parameter. Our approach applies 
to such physical systems as colloids interacting via a mildly repulsive potential. It can in 
principle, with moderate efforts, be extended to higher orders and to multipoint correlation 
functions.
\end{abstract}

\maketitle


When a suspension of polymer particles is cooled or compressed, a rapid slowing down 
of the dynamics occurs, and the suspension gradually becomes solid on experimental 
time scales, without any apparent change in structure \cite{PV86}. This colloidal glass transition is reminiscent of the phenomenology of 
molecular glasses. However colloids are conceptually simpler to analyze: the interaction 
potential often has a simple repulsive character (instead of a Lennard-Jones form) and their 
effective dynamics is Brownian (instead of Newtonian). They are also experimentally simpler 
to probe, since colloids are much larger than molecules in simple liquids.
An important class of colloids are those who interact via a bounded, repulsive potential. 
These, due to the existence of a finite energy scale in the potential, exhibit a 
re-entrant behaviour at high density --the glass melts upon increasing the density-- and the 
non-interacting liquid is recovered in the limit of infinite density \cite{LBH00,LLWL00,BMS10}. 
All particles evolve in a thermal bath (the solution) and thus undergo individual Brownian motions, 
while also interacting via a given pair-potential $v$.
To make our approach explicit, we chose to study the harmonic spheres model, where 
the pair-potential $v$ is taken to be of the form: 
$v(r) = \e \left(1 - \frac{r}{\s} \right)^2 \th \left(1-\frac{r}{\s} \right)$,
but most of the reasoning will be carried out for an arbitrary, sufficiently well-behaved fonction $v$.
This model was introduced by Durian~\cite{Du95} in the context of foam mechanics,
where $v(r)$ plays the role of an effective interaction potential that arises from a 
coarse-graining procedure,
but experimental realizations in colloids~\cite{SV99,CACGHA11} exist, and it became 
a model system to study glassy structure and dynamics~\cite{BW09b}.

The position $\vr_i(t)$ of each of the $N$ particles composing the colloidal suspension
evolve under Brownian dynamics, encoded in the following Langevin equations:
\begin{align}
& \frac{d \vr_i}{dt}(t) = - \sum_{j \ne i} \grad_{r_i} v \left( \vr_i(t) - \vr_j(t) \right) + \vec{\x}_i(t) , 
\label{langevin}
\end{align}
where $\vec{\xi}_i$ is a Gaussian white noise with variance $2T$ ($T$ is the bath temperature).
It is our goal to obtain quantitative predictions for the dynamics of the dense liquid phase of such 
colloidal suspensions upon approaching the glass transition. The only successful first-principles
theory to this day is the Mode-Coupling Theory (MCT) developped by G\"otze and 
collaborators~\cite{BGS84,Go99}. This is a closed, self-consistent equation of evolution for the 
relaxation of density fluctuations in equilibrium supercooled liquids which was initially 
 applied to particles evolving under Hamiltonian dynamics, but was later 
extended by Szamel and L\"owen~\cite{SL91} to interacting 
Brownian particles. No significant difference between these descriptions~\cite{FS05} emerges, 
at least within the MCT 
approximation. In both frameworks MCT predicts a strict dynamical arrest: below a 
critical temperature, 
density fluctuations are prevented from relaxing at long times, and ergodicity is spontaneously 
broken. While successes and failures of MCT are now well documented~\cite{RC05}, a 
systematic way of improving this approximation scheme to overcome the listed pitfalls is still 
lacking, since the original kinetic formulation of MCT involves physically motivated, but 
mathematically ill controlled approximations bearing on high-order correlation functions, 
and it contains no {\it a priori} small parameter.

The purpose of this letter is to present a new derivation of an MCT equation that bypasses several
known pitfalls at the same time: non-interacting particles ($v=0$) are exactly dealt with,
there exists a small dimensionless parameter, the strength of the potential $\e/T$, it follows from a well defined variational principle, it can easily be extrapolated 
to higher-orders,  and calculations for four-point quantities as well as for sheared systems 
can simply be implemented.

In order to gain insight into what could be a second order MCT, the idea of resorting to a field theoretic formulation is very appealing, since one can then exploit the standard toolbox of diagrammatic expansions and approximations 
developped in hard condensed matter and particle physics. Several crucial steps have been 
made over the past ten years in this direction. Preliminary works~\cite{DM86,RC05} have soon 
been shown to be inconsistent with micro-reversibility, a property which is not automatically 
conserved by standard approximations in field theory. MCT predicts an ergodic-nonergodic
transition and one must make sure it does not result
from a symmetry breaking approximation. Further attempts~\cite{MR05,ABL06,BR07} 
have considerably progressed into the conservation of micro-reversibility, but technical difficulties
led to either a non closed equation for density correlations, or to non physical behavior of the
solutions to the equations.

In recent years, Kawasaki and Kim~\cite{KK07b} obtained a result consistent with reversibility, 
which led to the same equation as that of the original MCT, but this result stems from a very cumbersome 
calculation, giving little hope of extending this result to higher orders. In the present letter, we suggest 
to further exploit the many-body theory tools used in condensed matter, by formally treating our
classical particles as bosons. We will see that this approach automatically solves several of the 
problems encountered in previous attempts of the formulation of a field-theoretic MCT, and provides
a transparent way to carry the approximations to next order, or to extend the calculation to 
different quantities, such as four-point correlators, or to non equilibrium settings, such as in 
sheared systems. We now proceed with a step-by-step presentation of our approach.

The $N$ coupled Langevin equations Eq.~(\ref{langevin}) can be described by a Fokker-Planck
equation governing the evolution of the probability $P(\{ \vr_i \} ,t)$ of finding each particle $i$ 
at a given position $\vr_i$ at a time $t$. As a consequence of micro-reversibility, the Fokker-Planck 
equation converges 
towards a Gibbsian equilibrium distribution $P(\{\vr_i\} ,t \to \io) = e^{-\sum_{i<j} v(\vr_i-\vr_j)/T}$.
A standard result~\cite{vanKampen} shows that the knowledge of the equilibrium distribution allows
one to render the Fokker-Planck operator Hermitian in the proper basis (this is 
often called the Darboux or supersymmetry transformation). Hermiticity
allows one to interpret this new Fokker-Planck equation as a quantum mechanical problem for 
interacting bosons. From there we use standard methods of quantum field 
theory~\cite{negele-orland} to describe the dynamics of the system, which is encoded in the 
following action:
\begin{align}
S[a,\ola] = & \int_{t,\vx} \left[ \ola ~ \partial_t  a + 
\vec{\nabla} \ola \cdot \vec{\nabla} a \right]  + V_{\text{eff}}[\ola a]/T .
\label{action}
\end{align}
which is expressed in terms of a pair of complex and conjugate fields $\ola$ and $a$. The 
kinetic term reflects the free diffusion of particles, and the two-body interactions with 
potential $v$ are expressed, in the quantum formulation, by the effective potential that now 
contains not only two but also three-body interactions as follows: 
\begin{align}
V_{\text{eff}}[\r] = & \frac{1}{4T} \int_{t,\vx,\vy,\vz} \!\!\!\!\!\! \r(\vx) \vec{\nabla}_{\vy} v \left( \vx - \vy \right) 
\r(\vy) \cdot \vec{\nabla}_{\vz} v \left( \vx - \vz \right) \r(\vz) \nonumber \\ 
& - \frac 1 2 \int_{t,\vx,\vy} \r(\vx) \D_{\vx} v \left( \vx - \vy \right) \r(\vy) ,
\label{effpot}
\end{align}
where $\r = \ola a$ is the physical density and $v$ is the pair potential between the colloids. 

It is important to notice that the symmetrization corresponds to a change of basis, 
so that this field theory does not represent directly the physical problem anymore. 
It was shown long ago~\cite{Ja79,JMS81} that the 
micro-reversibility of usual dynamical field-theories obtained from the regular Fokker-Planck 
equation is represented by a complicated, non linear transformation, making it very difficult
to preserve when performing mode-coupling approximations~\cite{ABL06}. In the symmetrized
theory, micro-reversibility is simply encoded in the hermiticity of the symmetrized operator, which 
is a symmetry easy to check and conserve even when performing approximations.
Furthermore, setting the pair-potential to $0$ cancels the effective potential in Eq.~(\ref{effpot}) 
and one recovers, without approximation, the free diffusion of colloids. Even if the 
dynamics described by the action Eq.~(\ref{action})  is only related to the real dynamics by a change 
of basis, careful analysis of the theory shows that far from initial and final conditions (in the ``bulk" of 
the time window), the difference between the modified 
dynamics and the real dynamics vanishes. Finally, keeping in mind that the pair potential $v(r)$ 
has an energy scale $\e$, 
we see that this approach gives a satisfactory basis for a perturbation expansion in powers 
of the dimensionless parameter $\e/T$. Our approach yields a theory that is expressed with ladder 
operators $a$ and $\ola$ that do not directly represent the physical density $\r = \ola a$. Introducing 
it by hand via a Lagrange multiplier, a field $\l$, we arrive at a field theory involving four 
independent fields, that we group into a single vector $\phi = (a,\ola,\l,\r)$. To conclude the layout for the diagrammatic expansion to come, we prefer 
working with fields defined by deviations around the saddle of the action, which describe a 
homogeneous and isotropic liquid state of mean density $\r_0$. Thus we set $\phi = (\sqrt{\r_0},\sqrt{\r_0},0,\r_0) + (\P,\olP,\l,\d \r)$, and obtain a four field field theory characterized by the following propagator:
\begin{align}
G_0^{-1}(k,\o) = \begin{pmatrix}
0 & i \o + k^2 & -\sqrt{\r_0} & 0 \\
-i \o + k^2 & 0 & -\sqrt{\r_0} & 0 \\
-\sqrt{\r_0} & -\sqrt{\r_0} & 0 & 1 \\
0 & 0 & 1 & u(k) \\
\end{pmatrix} , 
\end{align}
where $u(k) = \frac{k^2}{2 \r_0} \left[ \left(1+ \r_0 v(k)/T \right)^2 - 1 \right]$. The interaction part of the 
action is made of two cubic terms, one is $\int^*\g(k_1,k_2,k_3) \d \r(k_1) \d \r(k_2) \d \r(k_3)$ and the other is $\int^* \l(k_1) \P(k_2) \olP(k_3)$.
where $\g(k_1,k_2,k_3) = \frac 1{2T^2} \left[ k_1 \cdot k_2 ~ v(k_1) v(k_2) + perms. \right]$, and the symbol $\int^*$ stands for $\int_{k_1,k_2,k_3} \d(k_1+k_2+k_3)$.
We now turn to the procedure allowing us to determine the correlations of our fields.

The physical quantity that we are ultimately interested in is the matrix element of the two point 
correlator of the theory $G$ that describes density-density correlations. Since 
our theory contains four fields, $G$ is a $4\times 4$ matrix, with 10 independent entries.
Our goal is to obtain a closed equation bearing on the $\d \r$--$\d \r$ element only. Performing a 
double
Legendre transform of the dynamical partition function, one obtains a functional of the correlator
$G$, called the 2PI or the Luttinger-Ward functional, $\Phi[G]$. A careful analysis of this functional,
which can be found in modern field theory textbooks~\cite{cini,calzetta}, 
shows that it has several remarkable properties. It not only provides a variational principle to obtain the 
correlator $G$ (the functional attains its maximum when evaluated at the true correlator) but also gives access to the inverse of the correlator, since it is obtained as the functional 
derivative of $\Phi[G]$ with respect to $G$. Diagrammatically, it is composed of all two particle 
irreducible diagrams (2PI), allowing for simple truncations of the complete expression of $\Phi$.
Finally, any truncation of $\Phi$ 
can be shown to preserve the symmetries of the action, which we use  to 
conserve micro-reversibility when performing approximations. To obtain a self-consistent 
approximation for the two point correlator, one constructs an approximation
for the 2PI functional by selecting a certain subclass of diagrams that contribute to it. 
For example, the two simplest diagrams that contribute have the following topology: 
\begin{align}
\Phi[G] ~ = ~  \raisebox{-0.25cm}{\includegraphics[width=0.8cm]{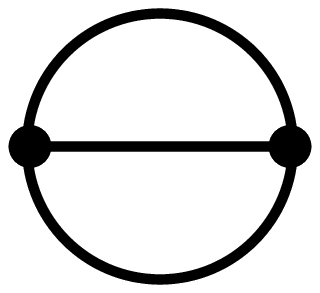}} ~ 
+ ~ \raisebox{-0.25cm}{\includegraphics[width=0.8cm]{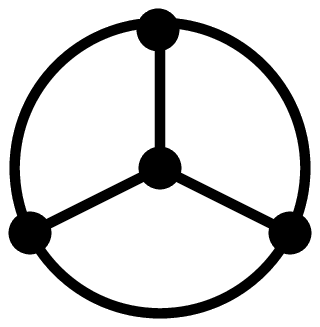}} ~ + ~ \ldots
\label{functional}
\end{align}
Then an expression for the vertex function $\Si$ (the inverse of the correlator) is obtained by 
functionally differentiating with respect to $G$.
Finally, a self consistent equation is obtained by exploiting the relationship that exists
between $\Si$ and $G$ (sometimes referred to as the Schwinger-Dyson equation): $\left( G_0^{-1} - \Si[G] \right) G = 1$.
This variational approach can be seen as the dynamical counterpart to the density functional theory 
of liquids.
 
So far, all these considerations are in principle exact. We now present the simplest 
self-consistent approximation that can be obtained within this formalism, and we will see that
we obtain an equation that has the exact same structure as that of the Mode-Coupling equation.
We now exploit that our theory contains a small parameter, namely the strength of the 
potential,
to select the lowest order beyond mean-field. We have two vertices in the theory, one of order 4 in $\e$, and one of order 0. We want to stay to the lowest non-trivial order, so 
 we can neglect the former.
 
 We only retain the simplest diagram in the expression of the 2PI functional 
 Eq.~(\ref{functional}), and will justify this {\it a posteriori}. We obtain an expression for the 
 vertex function 
 that can be inserted in the Schwinger-Dyson equation to yield:
 \begin{align}
 G_0^{-1} G(\vk,\t) \!\! = \!\!\! \int_{t',\vq} \!\!\!\!\!\! \G(\vk,\vq) G(\vk-\vq,\t-t') G(\vq,\t-t') G(\vk,t') 
 \label{schwinger}
 \end{align}
This is a matrix equation in which all correlators appear. 
Note that Eq.~(\ref{schwinger}) in itself already has the structure of the mode-coupling equation, in
which the memory kernel is a quadratic functional of the correlators, except that it applies to a 
matrix instead of a scalar. In order to write down an equation that involves the density-density 
correlator only, we must express all other correlators in terms of $C(\vk,t-t') = \la \d \r(-\vk,t') \d \r(\vk,t) \ra$. 
At the mean-field level, all 
correlators are proportional; we use these proportionality relations and insert them into 
Eq.~(\ref{schwinger}). The proportionality coefficients involve various powers of $\e/T$, and one then verifies that, when inserting the proportionality relations into the expression of the
2PI functional, all contributions coming from the diagrams that we neglected are indeed
of higher order in $\e/T$.

\begin{figure}[h]
\includegraphics[width=8cm]{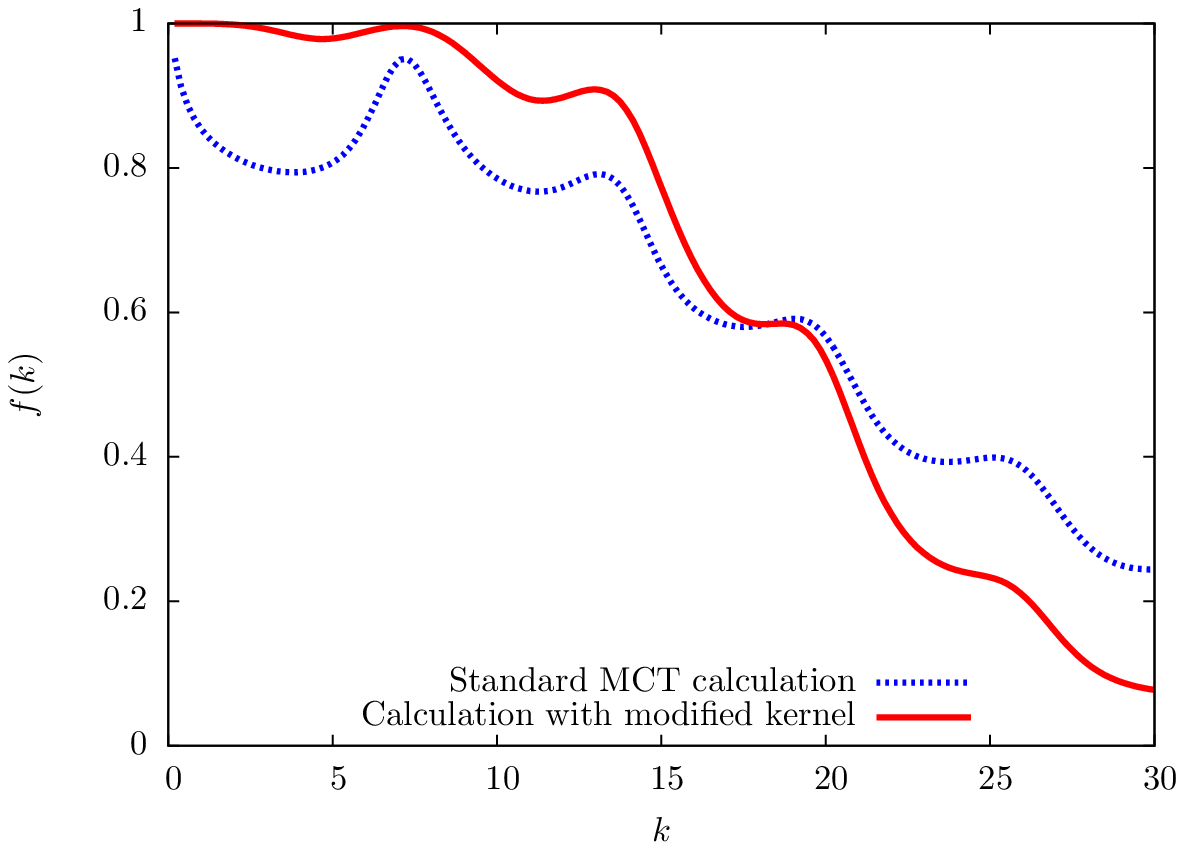}
\caption{Non-ergodicity parameter $f(k)$ calculated with the regular MCT kernel (full line) and 
within our modified kernel (dotted line), at packing fraction $0.53$ and temperature $10^{-4}$.}
\label{fig}
\end{figure}

In Eq.(\ref{functional}) only the topology of the diagrams is represented, but one has to draw all
possible diagrams from the vertices of the theory. Even for the simplest watermelon diagram, 
this involves 11 independent diagrams. Fortunately, to lowest order (order 2 in $\e/T$) only 
one diagram survives, and the final evolution equation for $C$ is:
\begin{align}
& 0 = -\partial_\t^2 C(\vk,\t) + \O(k)^2 C(\vk,\t) \label{MCTequation} \\
& \!\! + \!\! \frac 1 {2 \r_0} \!\! \int_{t',\vq} \!\!\!\!\!\! \MM(\vk,\vq) C(\vq,\t-t') C(\vk-\vq,\t-t') \partial_{t'} 
C(\vk,t') , \nonumber
\end{align}
where the memory kernel has the following expression:
\begin{equation}
\MM(\vk,\vq) = \left[ \vq^2 c(\vq) + (\vk-\vq)^2 c(\vk-\vq) \right]^2 
\label{final}
\end{equation}
Note that, as usual in field theoretic formulations, one has had to resort to a further approximation, i.e.
 setting $-v(\vk)/T = c(k)$. This  results from our treating the statics and the 
 dynamics on equal footing. At this order of approximation, this replacement is correct,
 as can be seen with a perturbation analysis of the equilibrium liquid, and allows for direct comparison
 with the regular MCT result.
The statics of equation \eqref{final} is closely similar to the original mode-coupling equation, apart 
from the slightly different wave-vector dependence that the original mode-coupling approach 
predicts, in which the factors $\vq^2$ and $(\vk-\vq)^2$ in the rhs of \eqref{final} are replaced 
with $\vk\cdot\vq$ and $\vk\cdot(\vk-\vq)$, respectively. Assuming that the density density correlation function does not
decay to zero at large times, one makes the usual ansatz: $\lim_{t \to \io} C(\vk,t) = \r_0 S(\vk) f(\vk)$, where $S(k)$
is the static structure factor related to the direct correlation function by $S(k)=1/(1-\r_0 c(k))$, and seek 
an equation for the non-ergodicity parameter $f(\vk)$. By Laplace transform methods 
one easily obtains:
\begin{equation}
\frac{f(k)}{1-f(k)} = \frac{\r_0 S(k)}{8 \p^2 k^4} \int_{\vq} \MM(\vk,\vq) S(\vq) S(\vk-\vq) f(\vq) f(\vk-\vq)
\label{MCTfk}
\end{equation}
We then numerically solve this equation with an iterative procedure. The only input is $c(k)$ for 
the equilibrium liquid, than can be calculated {\it e.g.} within the Hyper-Netted Chain approximation. Exactly as in the case of 
standard MCT, one finds that there exists a transition line $T_{MCT}(\r)$ above which $f(k)=0$ is the 
only solution, whereas below $T_{MCT}(\r)$, a nonzero $f(k)$ is found where ergodicity is 
spontaneously broken. {\it A posteriori}, we are inclined to view the MCT equation as a 
high-temperature expansion. 
In Fig.~\ref{fig} we show the resulting $f(k)$ at packing fraction $0.53$ and temperature $10^{-4}$.
The qualitative behavior is very similar, except for the limit $k \to 0$, where Eq.~\eqref{MCTfk}) is 
found to give $f(k) \underset{k \to 0}{\sim} 1$. The modification of the kernel implies that the 
absolute value of the transition temperature is slightly modified. For example at the packing 
fraction $0.53$, the regular MCT transition
is located approximately 
at $T_{MCT} \approx10^{-4}$, whereas our rough numerical estimate is $T_{MCT} \approx 8.10^{-4}$.

In this letter we have presented a comprehensive approach to write down mode-coupling equations
based on a variational principle. In the example of bounded interactions, we have shown that when
the strength of the interaction is taken as an expansion parameter, it is possible to write down, 
to lowest-order, a mode-coupling equation similar to the regular MCT equation. Our strategy can be extended in a variety of directions. The most obvious 
one is retaining higher orders in the expansion parameter $\e/T$. The resulting equation for the 
7-dimensional order parameter $G$ will pick up a $G^5$ contribution to its memory kernel. 
Retaining, after appropriate substitutions based on the leading order expansion \eqref{schwinger}, 
the next order in $\epsilon/T$ seems a tedious yet quite accessible task. It would also be of 
interest to examine whether qualitative differences show up if the full set of ten equations 
\eqref{schwinger} were solved. 
On our to-do list we also have more pressing wishes like implementing the 
so-called "thermodynamic of histories" formalism \cite{merolle,jack} and probing the relationships 
between ergodicity breaking and dynamic phase transitions. It would be interesting to investigate
sheared systems and compare our approach with existing extensions of MCT 
\cite{FC02}. The fate of ergodicity breaking and 
dynamic phase transitions under shear also belongs to our open questions.

We warmly acknowledge the advice and/or comments of G. Biroli, N. Dupuis, R.L. Jack, 
J.Kurchan, V. Lecomte, K. Miyazaki, G. Szamel and F. Zamponi. H. Jacquin acknowledges financial 
support from Capital Fund Management (CFM) foundation.


\end{document}